\documentclass[fleqn,twoside]{article}
\usepackage{espcrc2}

\usepackage{graphicx}
\usepackage[figuresright]{rotating}
\usepackage{psfig}

\newcommand{\HI}{H\,{\sc i}}

\newcommand{\la}{\ifmmode\stackrel{<}{_{\sim}}\else$\stackrel{<}{_{\sim}}$\fi} 
\newcommand{\ga}{\ifmmode\stackrel{>}{_{\sim}}\else$\stackrel{>}{_{\sim}}$\fi} 

\title{The Origin and Evolution of Cosmic Magnetism}

\author{B. M. Gaensler\address{Harvard-Smithonian Center for Astrophysics,
60 Garden Street MS-6, Cambridge, MA 02138, USA},
R. Beck\address{Max-Planck-Institut f\"ur Radioastronomie,
Auf dem H\"ugel 69, 53121 Bonn, Germany} and
L. Feretti\address{Istituto di Radioastronomia CNR/INAF, 
Via Gobetti 101, 40129 Bologna, Italy}}

\begin{document}

\begin{abstract}

Magnetism is one of the four fundamental forces.  However, the origin
of magnetic fields in stars, galaxies and clusters is an open problem in
astrophysics and fundamental physics.  When and how were the first fields
generated? Are present-day magnetic fields a result of dynamo action,
or do they represent persistent primordial magnetism? What role do
magnetic fields play in turbulence, cosmic ray acceleration and galaxy
formation? Here we demonstrate how the Square Kilometer Array (SKA)
can deliver new data which will directly address these currently unanswered
issues.  Much of what we present is based on an {\em all-sky survey of
rotation measures}, in which Faraday rotation towards $>10^7$ background
sources will provide a dense grid for probing magnetism in the Milky Way,
in nearby galaxies, and in distant galaxies, clusters and protogalaxies.
Using these data, we can map out the evolution of magnetised structures
from redshifts $z>3$ to the present, can distinguish between different
origins for seed magnetic fields in galaxies, and can develop a detailed
model of the magnetic field geometry of the intergalactic medium and of
the overall Universe. With the unprecedented capabilities of the SKA,
the window to the Magnetic Universe can finally be opened.

\end{abstract}

\maketitle

\section{Introduction}

Understanding the Universe is impossible without understanding magnetic
fields. They fill intracluster and interstellar space, affect the
evolution of galaxies and galaxy clusters, contribute significantly to
the total pressure of interstellar gas, are essential for the onset of
star formation, and control the density and distribution of cosmic rays
in the interstellar medium (ISM). But in spite of their importance, the
{\em evolution}, {\em structure}\ and {\em origin}\ of magnetic fields
are all still open problems in fundamental physics and astrophysics.
Specifically, we still do not know how magnetic fields are generated
and maintained, how magnetic fields evolve as galaxies evolve, what the
strength and structure of the magnetic field of the intergalactic medium
(IGM) might be, or whether fields in galaxies and clusters are primordial
or generated at later epochs. Ultimately, we would like to
establish whether there is a connection between magnetic field formation
and structure formation in the early Universe, and to obtain constraints
on when and how the first magnetic fields in the Universe were generated.

\begin{figure}
\centerline{\psfig{file=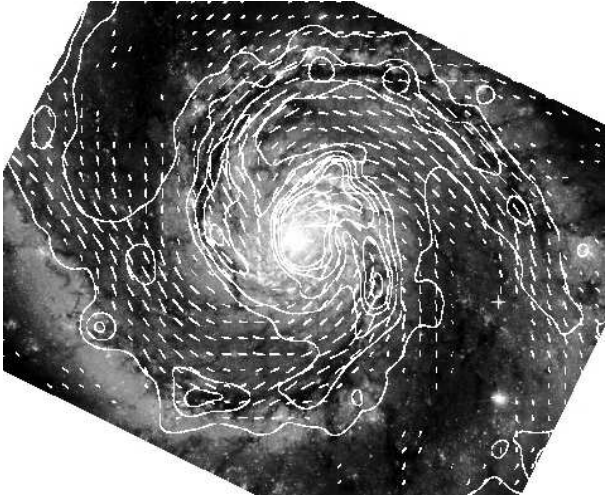,width=0.5\textwidth}}
\caption{The magnetic field of the grand design spiral galaxy M~51.
The image shows an optical {\em HST}\ image, overlaid with contours
showing the radio total intensity emission at 5~GHz.  The vectors show
the orientation of the magnetic field, as determined from 5~GHz linear
polarization measurements (Faraday rotation is small at this
frequency) \cite{fb04}.}
\label{fig_m51}
\end{figure}

\begin{figure}
\centerline{\psfig{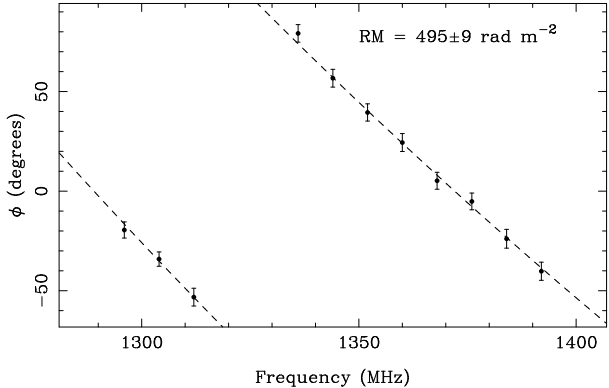}}
\caption{Spectropolarimetry of pulsar B1154--62, demonstrating
how the RM can be accurately determined using the position
angle of polarization in multiple channels within the
same observing band \cite{gmg98}.}
\label{fig_psr_rm}
\end{figure}

Most of what we know about astrophysical magnetic fields comes through
the detection of radio waves.  {\em Synchrotron emission}\ measures
the total field strength, while its {\em polarization}\ yields the 
orientation of the regular field
in the sky plane and also gives the field's degree of
ordering (see Figure~\ref{fig_m51}). Incorporating {\em Faraday rotation}\
yields a full three-dimensional view
by providing information on the
field component along the line of sight (see Figure~\ref{fig_psr_rm}),
while the {\em Zeeman effect}\ provides an independent measure of field
strength in cold gas clouds. However, measuring astrophysical magnetic
fields is a difficult topic still in its infancy, restricted to nearby
or bright objects.  In this Chapter we explain how the Square Kilometre
Array (SKA) can revolutionise the study of Cosmic Magnetism.

\section{Polarimetry with the SKA}

Much of what the SKA can contribute to our understanding of magnetic
fields will come from its {\em polarimetric capabilities}.  As discussed
in more detail elsewhere \cite{skabg,skafjh}, the main SKA specifications
which enable this capability will be {\bf high polarization purity}\
and {\bf spectropolarimetric capability}. The former will allow detection
of the relatively low linearly polarized fractions ($\la1$\%) from most
astrophysical sources seen down to sub-mJy levels, while the latter will
enable accurate measurements of Faraday rotation measures (RMs), intrinsic
polarization position angles and Zeeman splitting. Other minimum
requirements include a frequency coverage of 0.5--10~GHz, a field of view at 
1.4~GHz of 1~deg$^2$ which can be fully imaged at $1''$ resolution,
a significant ($\sim50\%$) concentration of the collecting area into a
central core of diameter $\sim5$~km, and a longest baseline of
$\sim300$~km.

It is important to appreciate how powerful the combination of high 
sensitivity, good polarization purity and full spectropolarimetry 
of the SKA will be.
With current instruments, the only way to simultaneously determine
accurate values for both the RM and intrinsic
polarization position angle is to make many observations across a broad
frequency range. This is not only time-consuming, but the analysis must
proceed cautiously, since various depolarizing effects have a strong
frequency dependence (e.g., \cite{sbs+98}).

The high sensitivity and broad bandwidth of the SKA eliminate these
difficulties: a single spectropolarimetric observation at a single
IF can simultaneously provide good estimates of both RM and position
angle, the limiting factor often only being the accuracy of ionospheric
corrections to the observed Faraday rotation.  For example, at an
observing frequency of 1.4~GHz with a fractional bandwidth of 25\%,
a 1-min SKA observation of a source with a linearly polarized surface
brightness of $\sim8$~$\mu$Jy~beam$^{-1}$ will yield a RM
determined to an accuracy $\Delta {\rm RM} \approx \pm 5$~rad~m$^{-2}$,
and its intrinsic position angle measured to within $\Delta \Theta
\approx \pm10^\circ$.  Measurements of this precision will be routinely
available in virtually any SKA observation of a polarized source.

\section{All-Sky Rotation Measures}
\label{sec_grid}

Currently $\sim$1200 extragalactic sources and $\sim$300 pulsars have
measured RMs. These data have proved useful probes of magnetic fields in
the Milky Way, in nearby galaxies, in clusters, and in distant Ly$\alpha$
absorbers. However, the sampling of such measurements over the sky
is very sparse, and most measurements are at high Galactic latitudes.

\begin{figure}
\centerline{\psfig{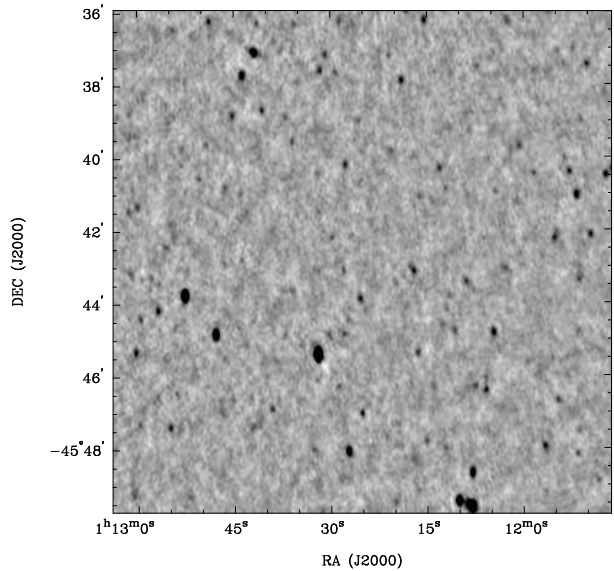}}
\caption{1.4~GHz ATCA image of a small section of the Phoenix Deep Field
\cite{hac+03}. The density of extragalactic 
sources in this total intensity image is
similar to what will be achievable in {\em polarized intensity}\ in a
1-hour 1.4~GHz observation with the SKA.}
\label{fig_phoenix}
\end{figure}

A key platform on which to base the SKA's studies of cosmic magnetism will
be to carry out an {\bf All-Sky RM Survey}, in which spectropolarimetric
continuum imaging of 10\,000~deg$^2$ of the sky can yield RMs for
approximately $2\times10^4$ pulsars and $2\times10^7$ compact polarized
extragalactic sources in about a year of observing time (see \cite{skabg}
for a detailed description). This data set will provide a grid of RMs at
a mean spacing of $\sim30'$ between pulsars and just $\sim90''$ between
extragalactic sources (see Figure~\ref{fig_phoenix}). 

In the following Sections, we describe some of the SKA experiments
which will provide new insight into the origin, structure and evolution
of cosmic magnetic fields.  Many of these observations stem directly
from the RM grid just described, or from deeper observations of specific
regions which will provide an even more closely spaced set of background
RMs. This discussion begins with the Milky Way and its ISM,
and then considers other sources at increasingly larger distances.

\section{The Local Universe}

\subsection{The Milky Way and Nearby Galaxies}  
\label{sec_nearby}

Key information on the processes which amplify and sustain large-scale
magnetic field structure in galaxies comes from characterising and
comparing the three-dimensional structure of the magnetic field in 
nearby galaxies.  Amongst the questions which we would like
to resolve are the number and location of magnetic reversals, the relation
between magnetic and optical spiral arms, the vertical magnetic structure
in the disk, the strength and structure of magnetic fields in galactic
halos, and the correspondence between magnetic field structure and other
parameters such as rotation speed, Hubble type and star formation rate.

In our Milky Way, the large sample of pulsar RMs obtained
with the SKA, combined with
distance estimates to these sources from parallax or from their
dispersion measures, can be inverted to yield a complete delineation
of the magnetic field in the spiral arms and disk on scales $\ga100$~pc
(e.g., \cite{sfss02}). Small-scale structure and turbulence can be probed
using {\em Faraday tomography}, in which foreground ionised gas produces
complicated frequency dependent Faraday features when viewed against
diffuse Galactic polarized radio emission (e.g., \cite{rbgd,rbhk1,rbul}).  
Magnetic field geometries in the Galactic halo and outer parts of the 
disk can be studied using the extragalactic RM
grid (see detailed discussion by \cite{skabg}).

In external galaxies, magnetic fields can be directly traced by diffuse
synchrotron emission and its polarization. For the nearest galaxies,
the SKA will provide sufficient resolution and sensitivity to carry out
Faraday tomography on this emission, allowing us to identify individual
features, structures and turbulent processes in the magnetoionised
ISM of these galaxies.
These measurements can be compared with
\HI, H$\alpha$ and continuum observations at a variety of wavelengths to
establish, in a much more direct way than is possible for the confused
lines of sight of our own Galaxy, how magnetic features interact with
other features in the ISM.

Synchrotron emission is seen only in magnetised regions in which either
cosmic ray electrons have been accelerated, or into which such
electrons have diffused. In other parts of a galaxy, particularly in
the outer disk and halo, we must rely on background RMs to map magnetic
fields. However, the current density of such sources is such that
only a handful of nearby galaxies can be studied in this way (e.g.,
\cite{rbhbb}). With the sensitivity of the SKA, deep observations
of nearby galaxies can provide $>10^5$ background RMs, and thus allow
fantastically detailed maps of the magnetic structure. Even at a distance
of 10~Mpc, most galaxies should be traversed by $\sim$50 sightlines
to polarized background sources. The sample of galaxies which could be
studied in this way is very large.

As discussed in detail by \cite{skabg}, the overall structure of galactic
magnetic fields provides a potential discriminant between dynamo and
primordial mechanisms for the origin and sustainment of these fields \cite{bbm}.
For example, the mean-field $\alpha$--$\Omega$ dynamo model makes specific
predictions for the symmetries seen in both azimuthal and vertical field
structure. Observations of background RMs and of diffuse polarized
synchrotron emission can allow direct measurements of the azimuthal
modes, but these measurements are limited at present to $\sim$20
galaxies \cite{b00}.  With the SKA, this sample can be increased by up
to three orders of magnitude.  These data can allow us to distinguish
between different conditions (e.g., strong density waves, high star
formation rate, or interactions) for excitation of various dynamo modes.
Meanwhile, the presence and prevalence of reversals in the disk field
structure, plus the structure of field in the halo, will together let
us distinguish between dynamo and primordial models for field origin in
different systems.

\subsection{Galaxy Clusters}

In clusters of galaxies, magnetic fields play a critical role in
regulating heat conduction \cite{cc98,nm01}, and may also both govern and
trace cluster formation and evolution. Estimates
of the overall magnetic field strength
come from inverse Compton detections in X-rays, from detection
of diffuse synchrotron emission,
from cold fronts and from simulations,
but our only direct measurements of field strengths and geometries come
from RMs of background sources (see \cite{skafjh}). Currently
just $\sim$1--5 such RM measurements can be made per cluster (e.g.,
\cite{gtd+01}); only by considering an ensemble of RMs averaged over
many systems can a crude picture of cluster magnetic field structures
be established \cite{ckb01}.

With the SKA, the RM grid can provide $\sim1000$ background RMs behind
a typical nearby cluster; a comparable number of RMs can be obtained 
for a more distant cluster through a deep targeted observation.
These data will allow us to derive a detailed map of the field in {\em
each}\ cluster.  With such information, careful comparisons of the
field properties in various types of cluster (e.g., those containing
cooling flows, those showing recent merger activity, etc.) at various
distances can be easily made. Furthermore, detailed comparisons between
RM distributions and X-ray images of clusters will become possible,
allowing us to relate the efficiency of thermal conduction to the
magnetic properties of different regions, and to directly study the
interplay between magnetic fields and hot gas.

Deep observations targeting individual clusters will also allow the
detection of low surface brightness synchrotron emission, allowing us
to explore the role of low-level magnetic fields, and providing vital
clues to the evolution of cluster magnetic fields.

\section{Galaxies at Intermediate Redshifts}

Many galaxies at intermediate redshifts ($0.1 \la z \la 2$)
are representative of the local population but at earlier epochs.
Measurements of the magnetic field in such systems thus provides direct
information on how magnetised structures evolve and amplify as galaxies
mature.  However, the linearly polarized emission from galaxies at these
distances will often be too faint to detect directly; Faraday rotation
thus holds the key to studying magnetism in these distant sources.

\subsection{Galaxies in the Foreground of Bright Extended Radio Sources}
\label{sec_fgd}

\begin{figure}
\centerline{\psfig{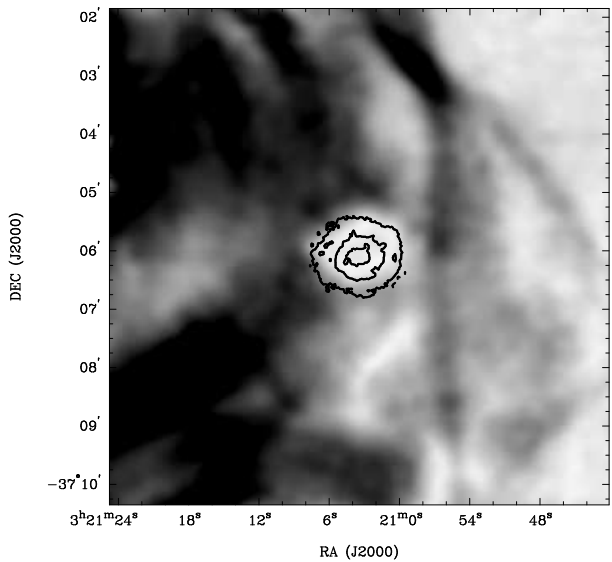}}
\caption{A 1.4~GHz VLA image of linearly polarized intensity from a small
region in the western lobe of Fornax~A, superimposed with optical contours
from the Digitized Sky Survey. The bright optical source corresponds
to the spiral galaxy NGC~1310, which produces strong depolarization
through its Faraday rotation towards the background polarized emission
from Fornax~A \cite{feve89}.}
\label{fig_fornax}
\end{figure}

At intermediate redshifts, we expect galaxies to be
$<1'$ in extent, 
too small to be adequately probed by the RM grid of compact background
sources discussed in \S\ref{sec_grid}.  However, there are many distant
{\em extended}\ polarized sources (e.g., many of the quasars and radio
galaxies in the 3C catalogue), which provide ideal background illumination
for probing Faraday rotation in galaxies which happen to lie along the
same line of sight.  These experiments can deliver maps of magnetic field
structures in galaxies more than 100 times more distant than discussed in
\S\ref{sec_nearby} above, and thus provide information on the evolution
of magnetic fields as a function of cosmic epoch.  Currently this
technique has been applicable in just a few fortuitous cases, such as for
NGC~1310 seen projected against a lobe of Fornax~A (Figure~\ref{fig_fornax};
\cite{feve89,sf92}), and for an absorption line line system at $z=0.395$
seen against PKS~1229--021 \cite{kpz92}.  With the greater sensitivity
of the SKA, many such systems lying in front of bright polarized radio
sources can be targeted. These objects are less evolved counterparts to
the nearby galaxies discussed in \S\ref{sec_nearby}; comparison of the
two populations will demonstrate how present
day magnetic fields emerge. For example, the $z=0.395$ system discussed
by Kronberg et al.\ \cite{kpz92} appears to be a spiral galaxy with a
bisymmetric magnetic field geometry and a magnetic field strength of
$\sim1-4$~$\mu$G, surprisingly similar to results from nearby galaxies.
Clearly a significant expansion of the
sample of such systems can provide strong constraints on the processes
which generate magnetic fields in galaxies.

\subsection{RM Statistics of Ly$\alpha$ Absorbers}

While the approach in \S\ref{sec_fgd} above allows us to study a
variety of individual systems in detail, an even larger sample,
suitable for statistical studies, can be accumulated by considering
unresolved sources in the foreground
of our RM grid.  Specifically, we expect that a large
number of the sources for which we measure
RMs will be quasars showing
foreground Ly$\alpha$ absorption; these absorption systems likely
represent the progenitors of present-day galaxies. If a large enough
sample of RMs for quasars at known redshift can be accumulated, a trend
of RM vs $z$ can potentially be identified. The form of this trend can
then be used to distinguish between RMs resulting from magnetic fields
in the quasars themselves and those
produced by fields in foreground absorbing clouds (e.g.,
\cite{wpk84}); detection of the latter effect would then directly trace
the evolution of magnetic field in galaxies and their progenitors.
This experiment has been attempted several times with
existing data sets, but only a relatively small number of absorbing
systems show a clear excess in RM. These data correspondingly provide
only marginal (if any) evidence for any evolution of RM with redshift
\cite{wpk84,wlo92,rbpw,ow95}. There are several limitations in
these studies, all of which can be circumvented using the SKA.

First, all quasar RM measurements are contaminated by the Galactic
contribution to the RM. The Galactic signal can be removed by considering
the RMs of neighboring background sources, but currently the sampling of
such nearby sources is extremely sparse, so that large uncertainties are
introduced through this foreground subtraction. The SKA's RM grid will
improve this situation by five orders of magnitude, allowing a complete
template for the Galactic foreground to be derived and then removed
(e.g., \cite{fsss01}).

Second, because {\em detected}\ RMs are lower than {\em intrinsic}\
RMs by a factor $(1+z)^2$, where $z$ is the redshift of the source in
which Faraday rotation is produced, an absorber at a redshift of $\sim3$
must have an RM signal $\sim80$ times the measurement uncertainty in
order to be considered significant. For many previous efforts, the
uncertainties introduced by removing the Galactic foreground are as
large as $\Delta{\rm RM} \sim 20-30$~rad~m$^{-2}$, making it difficult
to detect magnetic fields in distant objects unless the intrinsic RM
is extremely high.  With the SKA accurate RM measurements and precise
foreground subtraction will allow us to reliably probe magnetic field
evolution to much higher redshifts than has been previously possible.

Finally, an overriding limitation of previous studies has been the limited
number of sources per redshift bin. By correlating SKA RMs with new
surveys such as SDSS, a vast increase in the sample size is expected, so
that different evolutionary models can be clearly distinguished.

\section{Galaxies and Protogalaxies at $z \ga 2$}

At yet higher redshifts, we can take advantage of the sensitivity of
the deepest SKA fields, in which we expect to detect the synchrotron
emission from the youngest galaxies and proto-galaxies (see
contribution by Carilli, this volume).  The tight radio-infrared
correlation, which holds also for distant infrared-bright galaxies
\cite{cb01}, tells us that magnetic fields with strengths similar or
even larger than in nearby galaxies existed in some young objects, but
the origin of these fields is unknown.  The total intensity of
synchrotron emission can yield approximate estimates for the magnetic
field strength in these galaxies.  
between magnetic fields and cosmic rays.  In deep observations, the
SKA can detect and resolve spiral galaxies like M~51 out to redshifts
$z\approx2$, and to even larger distances if the star formation rate
is higher than in M~51 (see contribution by van der Hulst, this
volume).  Since a dynamo needs a few rotations or about $10^9$~yr to
build up a coherent field \cite{bp94}, detection of synchrotron
emission in young galaxies at high $z$ put constraints on the seed
field which potentially challenge dynamo models \cite{rbpw}.

Even very distant galaxies unresolvable by the SKA may reveal
polarized emission if the inclination is high or if the field
structure is asymmetric, e.g.\ due to tidal interaction or ram
pressure as has been observed in several galaxies (e.g., in NGC~2276;
\cite{rbhb}). Interactions were much more frequent in the early Universe
so excess polarization might be expected in some objects, especially if
magnetic fields were already present before compression.

\section{The IGM and Cosmic Field Geometry}
\label{sec_igm}

Fundamental to all the issues discussed above is the search for magnetic
fields in the IGM.  All of ``empty'' space may be magnetised, either by
outflows from galaxies, by relic lobes of radio galaxies, or as part of
the ``cosmic web'' structure.  Such a field has not yet been detected,
but its role as the likely seed field for galaxies and clusters, plus
the prospect that the IGM field might trace and regulate structure
formation in the early Universe, places considerable importance on its
discovery.  Once the magnetic field of the IGM is detected, a measurement
of the characteristic size scales of its fluctuations can allow us to
differentiate between the wide variety of mechanisms proposed for magnetic
field generation in the IGM (see \cite{kro94,gr01,ct02,wid02,gio03}
for extensive reviews).  While RMs of distant sources are potentially a
useful probe of the IGM, to date there has been no detection of magnetic
fields in the IGM; current upper limits on the strength of any such
field are model dependent, but suggest $|B_{\rm IGM}| \la 10^{-8}-10^{-9}$~G
\cite{kro94,bbo99}.

Using the SKA, this all-pervading cosmic magnetic field
can finally be identified through the RM grid proposed in
\S\ref{sec_grid}.  Just as the correlation function of galaxies yields
the power spectrum of matter, the analogous correlation function of this
RM distribution can provide the magnetic power spectrum of the IGM
as a function of cosmic epoch and over a wide range of spatial scales.
Such measurements will allow us to develop a detailed model of the
magnetic field geometry of the IGM and of the overall Universe.

As mentioned earlier, the observed RM is reduced by a factor $(1+z)^{2}$
over its intrinsic value. However, a simplistic assumption is that the
comoving magnetic flux and electron density are both constant, so that
electron density evolves as $(1+z)^3$, while the magnetic field evolves
as $(1+z)^2$ (e.g., \cite{wid02}). In this case, we expect the observed
RM to be proportional to $(1+z)^3$. While this is a naive calculation,
it demonstrates the general principle that RM signatures from distant
objects can be quite large, even if the foreground IGM has a relatively
small magnetic field.

\begin{figure}
\centerline{\psfig{file=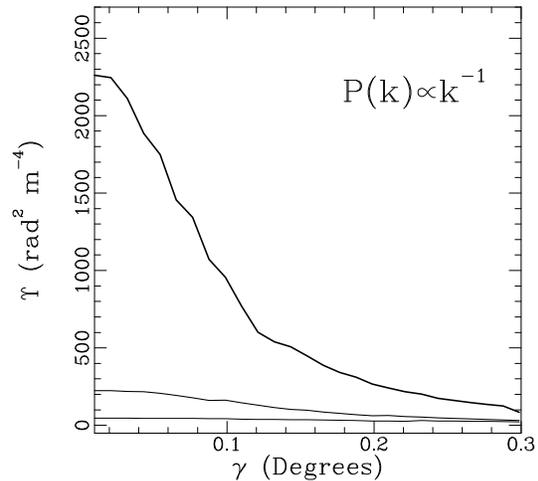,width=0.5\textwidth}}
\caption{RM correlation between pairs of background sources as a function
of separation angle, assuming a power spectrum for magnetic fluctuations
of the form $P_B(k) \propto k$ \cite{kol98}.  From bottom to top,
the three lines show
the correlation function for sources at redshifts of $z = 0.5, 1$ and 2.
Reproduced by permission of the AAS.}
\label{fig_kolatt}
\end{figure}

More robust treatments are presented by Kolatt \cite{kol98} and by
Blasi et al.\ \cite{bbo99}, who demonstrate in detail how the statistics
of RM measurements as a function of redshift can be used to remove the
foreground Galactic contribution, and then to extract the strength and power
spectrum of magnetic fields in the IGM.  These studies require a large
sample of RMs from sources at known redshift, which should be obtainable
by combining the SKA RM grid with the wide-field spectroscopic data bases
provided by SDSS and KAOS, plus with the all-sky multiband photometry
of LSST and SkyMapper.  For example, Kolatt \cite{kol98} shows that an
IGM field with strength $10^{-10}$~G can be detected with 5-$\sigma$
significance by considering extragalactic RM measurements over just
a few degrees of sky, provided that the density of RM measurements
is $\sim800$ sources~deg$^{-2}$; the resulting correlation function
is shown in Figure~\ref{fig_kolatt}.  Figure~2 of Beck
\& Gaensler \cite{skabg}
demonstrates that this is well beyond the capabilities of current
telescopes, but is easily achievable with the SKA. We note that
this experiment requires sub-arcsec spatial resolution at radio
wavelengths, so as to ensure accurate optical identification
of polarized radio sources.

\section{Primordial Fields ?}

As emphasised elsewhere in this volume, a large part of SKA  science
will be unanticipated and serendipitous discoveries, which do not easily
extrapolate from previous work. Thus, although speculative, we suggest
here some isolated experiments which might provide some constraints on
the strength and origin of magnetic fields at very early epochs.

\begin{figure}
\centerline{\psfig{file=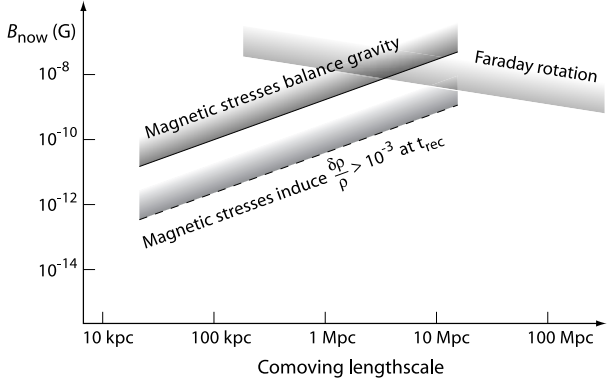,width=0.5\textwidth}}
\caption{Observational and theoretical
constraints on the magnetic field on various length scales 
at the time of recombination, $t_{rec}$.  A field is cosmogonically 
important if it can generate density perturbations of amplitude 
$\sim 10^{-3}$ at  $t_{rec}$, since these would have developed into 
gravitationally bound systems by the present time \cite{rees+00}.}
\label{fig_recomb}
\end{figure}

A magnetic field already present at the recombination era might
have affected the cosmogonic process \cite{sb+98}. The constraints
are summarised in Figure~\ref{fig_recomb}, in terms of the field's
characteristic length scale.

\begin{figure}
\centerline{\psfig{file=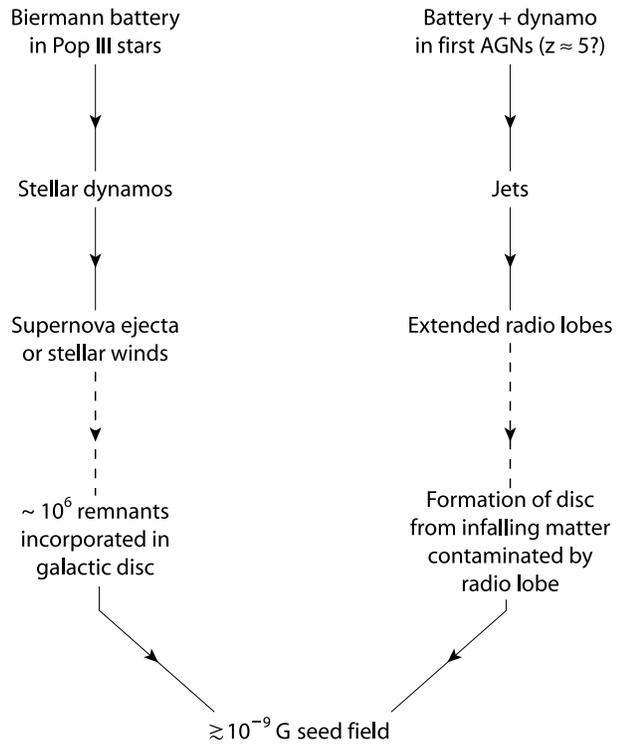,width=0.5\textwidth}}
\caption{Suggested mechanisms
through which Population~III stars and active galactic nuclei
can both produce seed magnetic fields of
strength $\ga10^{-9}$~G \cite{rees+04}.}
\label{fig_seed}
\end{figure}

A variety of cosmological processes can produce relatively high
primordial magnetic fields, of strength $10^{-10} - 10^{-9}$~G at $z\sim5$
(see \cite{gr01}).  Active galactic nuclei 
and violent star-formation
activity in young galaxies may be able to generate seed fields of similar
strength (e.g., \cite{crv94,klh99,va00,fl02}, as illustrated
schematically in Figure~\ref{fig_seed}. The dynamo mechanism may
then have amplified the seed field to the microgauss strength observed
in galaxies today. RM experiments with the SKA toward very high redshift
sources might provide evidence for such fields, taking advantage of the
RM~$\propto (1+z)^3$ effect discussed in \S\ref{sec_igm}.  For example,
we already know of gamma-ray bursts at redshifts as high as $z=4.5$
\cite{ahp+00}, and of radio galaxies at $z=5.2$ \cite{vds+99}.  With the
advent of projects such as {\em SWIFT}\ and LOFAR, it is virtually certain
that the sample of such high redshift sources will soon greatly expand.
With the high sensitivity of the SKA, linear polarization and Faraday
rotation should be detectable from some of these sources.  Using a
foreground template at lower redshift provided by the RM grid, the RM
contribution at high redshifts can then be isolated, yielding a possible
detection of magnetic fields at early epochs.

\section{Conclusion}

We have outlined the exciting new insights which the SKA
can provide into the origin, evolution and structure of cosmic magnetic
fields.  The sheer weight of RM statistics which the SKA can accumulate,
combined with deep polarimetric observations of individual sources, will
allow us to characterise the geometry and evolution of magnetic fields
in galaxies, in clusters and in the IGM from high redshifts through to the
present. We may also be able to 
provide the first constraints on when and how the first magnetic fields
in the Universe were generated.  Apart from these experiments which we can
conceive today, we also expect that the SKA will certainly discover new
magnetic phenomena beyond what we can currently predict or even imagine.

The main SKA specifications which enable us to reach our goals are high
polarization purity, spectropolarimetric capability, and a frequency
coverage of at least $0.5-10$~GHz.  The All-Sky RM Survey is best done
at a frequency of 1.4~GHz, perhaps simultaneously with other wide-field
continuum and \HI\ surveys.  Faraday tomography requires frequencies
$\sim0.5-1$~GHz for low-density, and $\sim1-5$~GHz for high-density
regions of the local ISM. Mapping of the polarized synchrotron emission in
the Milky Way, external galaxies and clusters is best done at frequencies
$\ga5-10$~GHz where Faraday depolarization is minimal. Other minimum
requirements are a field of view at 1.4~GHz of 1~deg$^2$ which can be fully
imaged at $1''$ resolution, a significant ($\sim50\%$) concentration
of the collecting area into a central core of diameter $\sim5$~km,
and a largest baseline of $\sim300$~km.

To summarise, incredibly rich data sets for the study of magnetic fields
await the advent of the SKA. With the unique sensitivity, resolution
and polarimetric capabilities of this next-generation radio telescope,
we can finally address fundamental questions relating to the physics of
cosmic magnetic fields, and their role in forming structure on all scales.

\vskip 0.2truein 

We thank JinLin Han, Martin Rees, Wolfgang Reich, Anvar Shukurov,
Dmitry Sokoloff, Kandu Subramanian, Marek Urbanik, Ira Wasserman
and Ellen Zweibel for useful comments.  B.M.G. acknowledges the
support of the National Science Foundation through grant AST-0307358.
The National Radio Astronomy Observatory is a facility of the National
Science Foundation operated under cooperative agreement by Associated
Universities, Inc. The Australia Telescope is funded by the Commonwealth
of Australia for operation as a National Facility managed by CSIRO.


\begin{thebibliography}{999}

\bibitem{ahp+00} Andersen, M.\,I.\, et al., 2000, A\&A, 364, L54

\bibitem{b00} Beck, R., 2000, Phil. Trans. R. Soc. Lond. A, 358, 777

\bibitem{skabg} Beck, R., Gaensler, B.\,M., this volume

\bibitem{bp94} Beck, R., Poezd, A.\,D., Shukurov, A., Sokoloff, D.,
1994, A\&A, 289, 94

\bibitem{bbm} Beck, R., Brandenburg, A., Moss, D., Shukurov, A., Sokoloff, D.,
1996, ARA\&A, 34, 155

\bibitem{bbo99} Blasi, P., Burles, S., Olinto, A.\,V., 1999,
ApJ, 514, L79

\bibitem{cb01} Carilli, C.\,L., Bertoldi, F., Omont, A., Cox, P.,
McMahon, R.\,G., Isaak, K.\,G., 2001, AJ, 122, 1679

\bibitem{ct02} Carilli, C.\,L., Taylor, G.\,B., 2002, ARA\&A, 40, 319

\bibitem{crv94} Chakrabarti, S.\,K., Rosner, R., Vainshtein, S.\,I.,
1994, Nature, 368, 434

\bibitem{cc98} Chandran, B.\,D.\,G., Cowley, S.\,C., 1998, Phys.\ Rev.\
Lett., 80, 3077

\bibitem{ckb01} Clarke, T.\,E., Kronberg, P.\,P., B\"{o}hringer, H.,
2001, ApJ, 547, L111

\bibitem{skafjh} Feretti, L., Johnston-Hollitt, M., this volume

\bibitem{fb04} Fletcher, A., Beck, R., et al., 2004, in preparation

\bibitem{feve89} Fomalont, E.\,B., Ebneter, K.\,A., van Bruegel,
W.\,J.\,M., Ekers, R.\,D., 1989, ApJ, 346, L17

\bibitem{fsss01} Frick, P., Stepanov, R., Shukurov, A., Sokoloff, D.,
2001, MNRAS, 325, 649

\bibitem{fl02} Furlanetto, S.\,R., Loeb, A., 2001, ApJ, 556, 619

\bibitem{rbgd} Gaensler, B.\,M., Dickey, J.\,M., McClure-Griffiths,
N.\,M., Green, A.\,J., Wieringa, M.\,H., Haynes, R.\,F., 2001, ApJ,
549, 959

\bibitem{gmg98} Gaensler, B.\,M., Manchester, R.\,N., Green, A.\,J.,
1998, MNRAS, 296, 813

\bibitem{gio03} Giovannini, M., 2003, preprint (astro-ph/0312614)

\bibitem{gtd+01} Govoni, F., Taylor, G.\,B., Dallacassa, D.,
Feretti, L., Giovannini, G., 2001, A\&A, 379, 807

\bibitem{gr01} Grasso, D., Rubinstein, H.\,R., 2001, Phys.\ Rep., 348, 163

\bibitem{rbhbb} Han, J.\,L., Beck, R., Berkhuijsen, E.\,M., 1998, A\&A,
335, 1117

\bibitem{rbhk1} Haverkorn, M., Katgert, P., de Bruyn, A.\,G., 2003,
A\&A, 403, 1031

\bibitem{hac+03} Hopkins, A.\,M., Afonso, J., Chan, B., Cram, L.\,E., 
Georgakakis, A., Mobasher, B., 2003, AJ, 125, 465

\bibitem{rbhb} Hummel, E., Beck, R., 1995, A\&A, 303, 691

\bibitem{kol98} Kolatt, T., 1998, ApJ, 495, 564

\bibitem{kro94} Kronberg, P.\,P., 1994, Rep.\ Prog.\ Phys., 57, 325

\bibitem{klh99} Kronberg, P.\,P., Lesch, H., Hopp, U., 1999, ApJ, 511, 56

\bibitem{kpz92} Kronberg, P.\,P., Perry, J.\,J., Zukowski, E.\,L.\,H.,
1992, ApJ, 387, 528

\bibitem{nm01} Narayan, R., Medvedev, M.\,V., 2001, ApJ, 562, L129

\bibitem{ow95} Oren, A.\,L., Wolfe, A.\,M., 1995, ApJ, 445, 624

\bibitem{rbpw} Perry, J.\,J., Watson, A.\,M., Kronberg, P.\,P., 1993,
ApJ, 406, 407

\bibitem{rees+00} Rees, M., {\it New Perspectives in Astrophysical Cosmology},
2nd edition, 2000, Cambridge Univ. Press

\bibitem{rees+04} Rees, M., in {\it Magnetic Fields in the Universe},
ed. R~Wielebinski, 2004, Springer, Berlin

\bibitem{sf92} Schulman, E., Fomalont, E.\,B., 1992, AJ, 103, 1138

\bibitem{sbs+98} Sokoloff, D.\,D., Bykov, A.\,A., Shukurov, A.,
Berkhuijsen, E.\ M., Beck, R., Poezd, A.\,D., 1998, MNRAS, 
299, 189

\bibitem{sfss02} Stepanov, R., Frick, P., Shukurov, A., Sokoloff, D.,
2002, A\&A, 391, 361

\bibitem{sb+98} Subramanian, K., Barrow, J.\,D., 1998, Phys. Rev. Lett., 81, 3575

\bibitem{rbul} Uyan{\i}ker, B., Landecker, T.\,L., Gray A.\,D., Kothes, R.,
2003, ApJ, 585, 785

\bibitem{va00} V\"{o}lk, H.\,J., Atoyan, A.\,M., 2000, ApJ, 541, 88

\bibitem{vds+99} van Breugel, W., De Breuck, C., Stanford, S.\,A., Stern,
D., R\"{o}ttgering, H., Miley, G., 1999, ApJL, 518, L61

\bibitem{wpk84} Welter, G.\,L., Perry, J.\,J., Kronberg, P.\,P.,
1984, ApJ, 279, 19

\bibitem{wid02} Widrow, L.\,M., 2002, Rev.\ Mod.\ Phys., 74, 775

\bibitem{wlo92} Wolfe, A.\,M., Lanzetta, K.\,M., Oren, A.\,L.,
1992, ApJ, 388, 17

\end{thebibliography}
\end{document}